\documentclass[9pt,conference]{IEEEtran}
\IEEEoverridecommandlockouts
\usepackage{cite}
\usepackage{amsmath,amssymb,amsfonts}
\usepackage{algorithmic}
\usepackage{graphicx}
\usepackage{textcomp}
\usepackage{xcolor}

\usepackage{amssymb}
\usepackage[hidelinks]{hyperref}
\usepackage{multirow}
\usepackage{makecell}
\usepackage{xcolor}
\usepackage{booktabs}
\usepackage{tikz}
\usepackage{hyperref}
\usepackage[labelfont=small]{caption}

\usepackage{titlesec}
\titlespacing*{\section}
{0pt}{1.0ex plus 1ex minus .2ex}{0.3ex plus .2ex}
\titlespacing*{\subsection}
{0pt}{0.3ex plus 1ex minus .2ex}{0.2ex plus .2ex}

\def\BibTeX{{\rm B\kern-.05em{\sc i\kern-.025em b}\kern-.08em
    T\kern-.1667em\lower.7ex\hbox{E}\kern-.125emX}}

\usepackage[margin=0.75in]{geometry} % Adjust Margins
\begin{document}

\title{\textbf{SELMA}: A \textbf{S}peech-\textbf{E}nabled \textbf{L}anguage \textbf{M}odel for Virtual \textbf{A}ssistant Interactions}
%\title{An Audio Language Model for Multi-Task Virtual Assistant Interactions}

\author{
    \IEEEauthorblockN{Dominik Wagner\IEEEauthorrefmark{2}\IEEEauthorrefmark{1}\thanks{\IEEEauthorrefmark{1}Work done during an internship at Apple.}, Alexander Churchill\IEEEauthorrefmark{3}, Siddharth Sigtia\IEEEauthorrefmark{3}, Erik Marchi\IEEEauthorrefmark{3}}
    \IEEEauthorblockA{\IEEEauthorrefmark{2}TH Nürnberg, \IEEEauthorrefmark{3}Apple}
    \IEEEauthorblockA{\texttt{\href{mailto:alex.churchill@apple.com}{alex.churchill@apple.com}}}
}

% \author{\IEEEauthorblockN{Dominik Wagner}
% \IEEEauthorblockA{\textit{TH Nuernberg}
% %\texttt{\href{mailto:dominik.wagner@th-nuernberg.de}{dominik.wagner@th-nuernberg.de}}
% }
% \and
% \IEEEauthorblockN{Alexander Churchill}
% \IEEEauthorblockA{\textit{Apple}
% %\texttt{\href{mailto:alex.churchill@apple.com}{alex.churchill@apple.com}}
% }
% \and
% \IEEEauthorblockN{Siddarth Sigtia}
% \IEEEauthorblockA{\textit{Apple}
% %\texttt{\href{mailto:sidsigtia@apple.com}{sidsigtia@apple.com}}
% }
% \and
% \IEEEauthorblockN{Erik Marchi}
% \IEEEauthorblockA{\textit{Apple}
% %\texttt{\href{mailto:emarchi@apple.com}{emarchi@apple.com}}
% }
% % \and
% % \IEEEauthorblockN{5\textsuperscript{th} Given Name Surname}
% % \IEEEauthorblockA{\textit{dept. name of organization (of Aff.)} \\
% % \textit{name of organization (of Aff.)}\\
% % City, Country \\
% % email address or ORCID}
% % \and
% % \IEEEauthorblockN{6\textsuperscript{th} Given Name Surname}
% % \IEEEauthorblockA{\textit{dept. name of organization (of Aff.)} \\
% % \textit{name of organization (of Aff.)}\\
% % City, Country \\
% % email address or ORCID}
% }

\definecolor{nodefill}{RGB}{213, 232, 212}
\definecolor{nodedraw}{RGB}{130,179,102}
\newcommand*\circled[1]{\tikz[baseline=(char.base)]{
            \node[fill=nodefill,shape=circle,draw=nodedraw,inner sep=0.3pt] (char) {#1};}
            }

\let\OLDthebibliography\thebibliography
\renewcommand\thebibliography[1]{
  \OLDthebibliography{#1}
  \setlength{\parskip}{0pt}
  \setlength{\itemsep}{0pt plus 0.3ex}
}

\maketitle

\begin{abstract}
In this work, we present and evaluate SELMA, a Speech-Enabled Language Model for virtual Assistant interactions that integrates audio and text as inputs to a Large Language Model (LLM). 
SELMA is designed to handle three primary and two auxiliary tasks related to interactions with virtual assistants simultaneously within a single end-to-end model.
We employ low-rank adaptation modules for parameter-efficient training of both the audio encoder and the LLM.  
Additionally, we implement a feature pooling strategy enabling the system to recognize global patterns and improve accuracy on tasks less reliant on individual sequence elements. 
Experimental results on Voice Trigger (VT) detection, Device-Directed Speech Detection (DDSD), and Automatic Speech Recognition (ASR), demonstrate that our approach both simplifies the typical input processing pipeline of virtual assistants significantly and also improves performance compared to dedicated models for each individual task. 
SELMA yields relative Equal-Error Rate improvements of 64\% on the VT detection task, and 22\% on DDSD, while also achieving word error rates close to the baseline. 
% Calculation: 
% DDSD: Best baseline: GPT2-XL + Whisper: 10 - 7.78 / 10 = 22.2% 
% VT:   Best baseline: UAD: 0.33 - 0.12 / 0.33 = 63.6%
% By combining acoustic and lexical features, our system demonstrates significant improvements in error rates and overall task accuracy. 
%This work highlights the potential of integrating multimodal inputs into a single LLM for comprehensive virtual assistant capabilities to enable more integrated and efficient processing of user interactions.
\end{abstract}


\section{Introduction}\label{sec:intro}
\begin{figure}[t]
\centering
\includegraphics[width=0.5\textwidth]{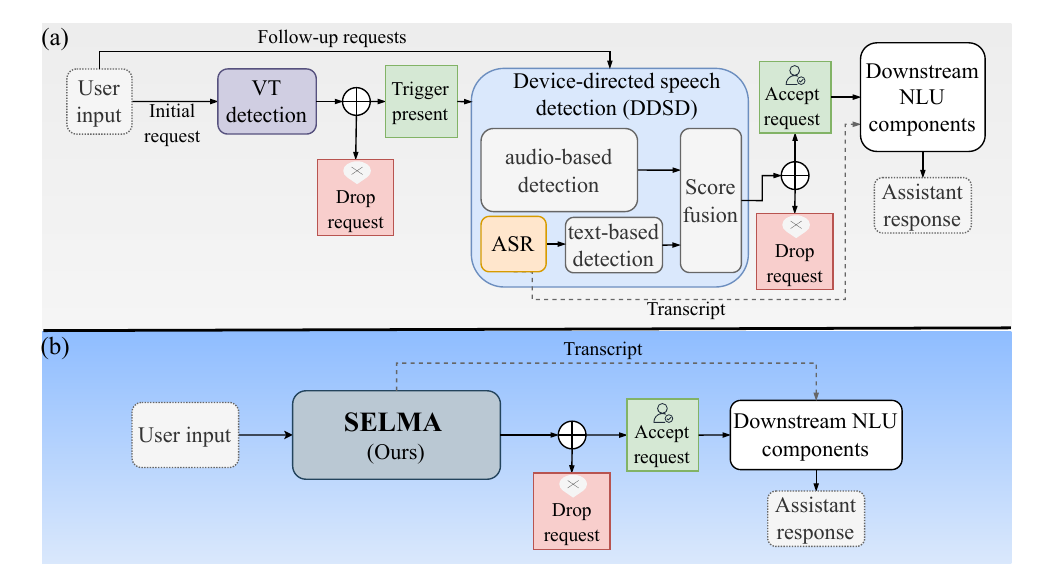}
\vspace{-8mm}
\caption{A typical pipeline for user input processing in virtual assistants (a) and our proposed approach (b).   
}
\label{fig:pipeline}
\vspace{-4mm}
\end{figure}

Voice-activated virtual assistants enable users to engage with smartphones, smartwatches, augmented reality headsets, speakers, and earphones through spoken commands. 
Typical pipelines for processing interactions with virtual assistants, such as the one depicted in Figure~\ref{fig:pipeline}~(a), involve multiple models, each specializing in specific tasks. 
% Usually, a trigger phrase or the press of a button precedes the first command, --to distinguish audio that is directed towards the device-- from background speech \cite{voicetrigger23}. 
A trigger phrase or button press usually precedes the first command, distinguishing speech intended for the device from background noise \cite{voicetrigger23}.
The problem of detecting a trigger phrase is referred to as voice trigger (VT) detection \cite{sigtia18_interspeech,higuchi2024multichannelvoicetriggerdetection}, wake-word detection \cite{jose20_interspeech,ghosh22_interspeech}, or keyword spotting \cite{sainath15kws,michaely17kws,ng23kws}. 
Subsequent interactions with the virtual assistant may not necessarily require the inclusion of a trigger phrase and are therefore processed with device-directed speech detection (DDSD) components using longer acoustic and lexical context \cite{wagner2024multimodal}. 
DDSD is concerned with determining whether a virtual assistant is being addressed, without the requirement of a trigger phrase preceding each voice command \cite{shriberg12_interspeech,mallidi18_interspeech,garg22_interspeech}. 
This task is more complex than voice trigger detection, due to the potentially missing trigger phrase indicating the start of a voice command. 
DDSD is also used as an additional mechanism to mitigate non-trigger speech that may have been falsely identified as the trigger phrase by the VT system, due to background noise or speech that sounds similar to the trigger phrase \cite{garg21_interspeech}. 
% Using DDSD as an additional false trigger mitigation mechanism allows for smaller and more efficient VT models
% TODO: Say sth like Typically, the input signal is broken down into audio and text and different systems process audio and text 
Typically, the input signal is broken down into audio and text, and separate DDSD systems are required to either operate on lexical feature \cite{dighe2020lrnn,jeon2019lrnn} (text-based DDSD) or audio feature \cite{norouzian2019ddsd,rudovic23sdsd} (audio-based DDSD) level, followed by a late fusion scheme to generate the final directedness decision \cite{huang19i_interspeech}. 
%Models operating only on textual features are referred to as Out-of-Domain Language Detection (ODLD) systems \cite{zheng2020odld,yin2022ood}. 
DDSD systems that operate on lexical features alone require an additional upstream automatic speech recognition (ASR) component to generate input features such as 1-best hypotheses or decoding lattices. 
Furthermore, ASR transcripts typically provide the input to downstream Natural Language Understanding (NLU) components, which act on the user input and generate the assistant's response. 
More recently, multimodal systems have been employed to jointly consider audio, text and visual modalities for the DDSD task \cite{krishna2023modalitydropoutmultimodaldevice,palaskar2024multimodallargelanguagemodels,wagner2024multimodal}. 
%Although VT detection and DDSD algorithms are generally precise and reliable, the operating point may allow non-trigger speech or background noise to falsely trigger the device, despite the user not having spoken the trigger phrase. 
% TODO: Simplify this paragraph: Given an input signal, it comprises the audio and the text transcription obtained from an ASR system and typically these modalities are dealt with different systems. 
%ASR transcripts obtained from the audio can be used as an additional signal in a solely text-based out-of-domain language detection task to gain an additional signal about the user's intent.  
%Downstream components such as dictation applications or web-based searches also operate on ASR transcripts. 
% A typical pipeline for processing user inputs in virtual assistants is depicted in Figure~\ref{fig:pipeline}. 

%%%%%%%%%%%% DIALOG ACT %%%%%%%%%%%
% An additional signal closely related to DDSD is dialogue act (DA) classification \cite{core1997coding,stolcke2000dialogue,grau2004da}. 
% DA classification involves categorizing querys based on the roles they fulfill within a conversation, such as questions, commands or statements. 
% Knowing the DA label, can help to better distinguish whether an query is intended for the virtual assistant.  
%and to direct the user input to the correct NLU component for downstream processing. 
Pretrained Large Language Models (LLMs) are increasingly deployed on devices, such as smartphones, serving as foundational backbones that can be directly used or finetuned to perform new tasks \cite{chen2023speedneedondeviceacceleration,gunter2024appleintelligencefoundationlanguage,abdin2024phi3technicalreporthighly}. 
While LLMs exhibit strong in-context learning capabilities, allowing them to perform new tasks without additional training \cite{kojima2022zeroshot,brown2020gpt3}, their implementation for tasks such as VT detection and DDSD is challenging. 
Firstly, in real-world applications, relying solely on text generated by ASR systems can suffer from errors due to audio distortions or noisy environments. 
Secondly, even under controlled conditions, relying solely on transcribed text without additional acoustic cues can result in inaccurate decisions. 
Numerous audio-enhanced LLMs such as Audio Flamingo \cite{kong2024audioflamingo}, SALMONN \cite{tang2024salmonn}, Qwen-Audio \cite{chu2023qwenaudio}, AudioPaLM \cite{rubenstein2023audiopalm}, and Pengi \cite{deshmukh2023pengi} have been proposed to bridge the gap between audio and text modalities. 
These models are usually equipped with an audio encoder that generates audio representations, which are then mapped into the latent space of the LLM. 
Audio LLMs have demonstrated state-of-the-art performance on audio captioning, audio question-answering, and various classification tasks for audio events, acoustic scenes, and music genres. 
%Motivated by these results, we aim to further enhance the functionality of Audio LLMs by training a model that integrates multimodal input data and enables comprehensive multi-task decision-making in the context of virtual assistants.  

% In a typical processing pipeline for interactions with virtual assistants, such as the one depicted in Figure~\ref{fig:pipeline}, multiple models are involved, each specialized in handling specific tasks. 
In this work, we aim to reduce the complexity of the pipeline depicted in Figure~\ref{fig:pipeline}(a) and explore the feasibility of replacing it with a single model capable of understanding the tasks related to processing virtual assistant interactions (cf. Figure~\ref{fig:pipeline}(b)). 
We leverage the strengths of LLMs, such as their ability to generalize across diverse tasks, their capacity to understand and generate coherent and contextually appropriate language, and their effectiveness in learning from diverse large-scale datasets. 
These advantages allow our model to perform well in scenarios that previously required separate, specialized models, thus streamlining the processing pipeline while maintaining or even improving performance.
Specifically, we explore a Speech-Enabled Language Model for virtual Assistant interactions (SELMA) that unifies five tasks crucial for processing user inputs during interactions with a virtual assistant. 
% (3) ODLD, (4) DA classification, a
SELMA is capable of jointly performing the three primary tasks VT detection, ASR, and DDSD, as well as the two auxiliary tasks text-based DDSD and Dialog Act (DA) \cite{core1997coding} classification.
The model's objective is not only to learn each task individually from the streaming audio captured by the device's microphone, but also being able to perform multiple tasks in sequence (e.g. first ASR followed by VT detection) based on a single input query. 
%Additionally, we employ ODLD and Dialog Act (DA) \cite{core1997coding} classification as auxiliary tasks serving as regularizers. 

Our work differs from previous attempts \cite{wagner2023multimodal,wagner2024multimodal,palaskar2024multimodallargelanguagemodels} to equip LLMs with non-lexical modalities in the context of virtual assistants in three major ways: 
% TODO: combine (2) and (3): Compared to previous works we present a multimodal end-to-end system where the audio encoder and the LLM are jointly trained. 
% Other studies compute an averages of all audio features
% (2) other modalities, such as audio, are generated with separate models and are then added as input features to the LLM, whereas the audio encoder in our model is a native part of the system,
(1) Previous models concerned with tasks such as DDSD focus exclusively on a single task, whereas our model jointly considers multiple tasks, (2) encoder models for non-lexical modalities are commonly frozen, whereas we jointly optimize the audio encoder and the LLM, and (3) other studies reduce the length non-lexical modalities, i.e., non-lexical inputs are represented by a single vector aggregated over the sequence length, whereas our model considers a variable-length sequence of audio representations, thereby keeping its ability to perform ASR and the ability to attend to specific regions in the query. 
% TODO: NOTE: Maybe move this to introduction: 

Additionally, we implement a feature pooling strategy to enhance performance on tasks that benefit from aggregated information rather than sequential data. 
This approach enables the system to recognize global patterns and improves accuracy on tasks less reliant on individual sequence elements. 
Our proposed system outperforms strong baselines specifically trained for VT detection and DDSD, while also performing well on ASR. 

% TODO: END - NOTE: Maybe move this to introduction
% The primary motivation for this research stems from the need to create models that can process multimodal inputs and perform multiple tasks related to virtual assistants. 
% Specifically, we seek to develop a model that can perform tasks such as detecting trigger phrases, identifying dialog acts, and determining whether an query is intended for a virtual assistant. These capabilities are essential for improving the interactivity and usefulness of virtual assistants in real-world scenarios.
% By developing a model that can listen to audio, read textual prompts, form decisions based on contextual understanding, and reason about these decisions, we aim to unify and improve upon tasks related to a virtual assistant. 

% We aim to further enhance the functionality of Audio LLMs by training a model that integrates multimodal inputs, enabling comprehensive decision-making and reasoning in the context of virtual assistants. 
% Our goal is to enable ASR, voice trigger detection and ODLD tasks, as well as SDSD, which is more complex the model's decision-making and reasoning abilities are crucial. 
% We demonstrate its ability to interpret audio signals, integrate them with textual data, and provide coherent, context-aware responses. 
% By advancing the multimodal processing capabilities of LLMs, we aim to enhance the functionality and reliability of virtual assistants, making them more adept at understanding and responding to user inputs. 
\section{Method}

\subsection{Multimodal LLM}
SELMA is built on top of the instruction-finetuned version of Qwen-Audio \cite{chu2023qwenaudio} known as Qwen-Audio-Chat. 
We chose this architecture, because it has been shown state-of-the-art results on a variety of tasks closely related to ours (e.g. keyword spotting and ASR). 
Qwen-Audio consists of two main components: an audio encoder and a language model. 
The audio encoder is responsible for generating latent representations from an audio sequence, which are subsequently processed by an LLM. 
The audio encoder is a Transformer \cite{vaswani17attention} with 32 layers based on the largest version of Whisper (\texttt{whisper-large-v2}) \cite{radford2022robust}, and has $\sim$650M parameters. 
Audio representations are inserted at audio placeholder token positions, which are then processed by the language model. 
The LLM backbone is based on Qwen \cite{bai2023qwentechnicalreport} ($\sim$7.4B parameters) and takes both audio context in the form of processed audio tokens and textual context as inputs. 

\subsection{Our Approach}
\begin{figure}[t]
\centering
\includegraphics[width=0.58\textwidth]{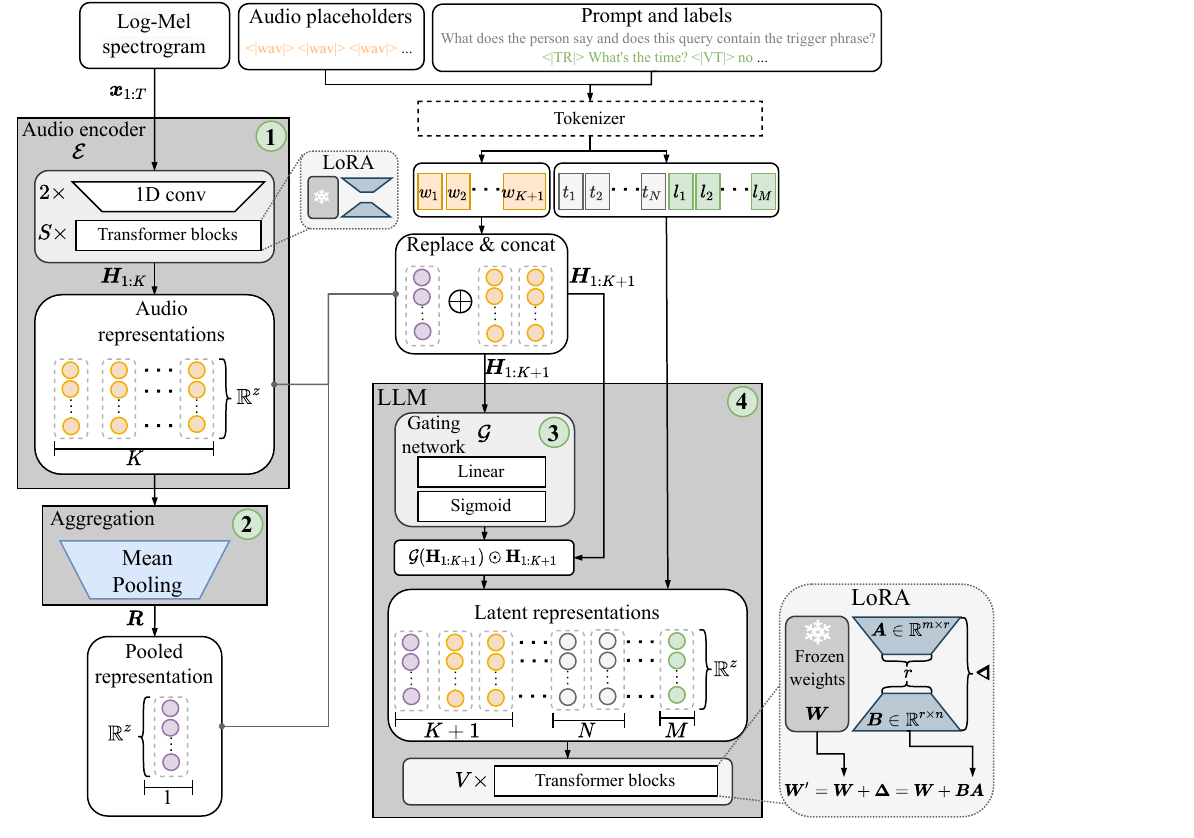}
\vspace{-6mm}
\caption{Overview of our approach.  
}
\label{fig:model}
\vspace{-6mm}
\end{figure}
Figure~\ref{fig:model} illustrates SELMA, which consists of four main components:  
The first main component \circled{\textbf{1}} is an audio encoder $\mathcal{E}$, which generates a sequence of latent audio representations from an input sequence given by log-Mel spectrogram features of length $T$. 
$\mathcal{E}$ consists of 1D convolutional layers responsible for downsampling the spectral features followed by $S$ transformer blocks. 
The result is a high-dimensional tensor, denoted as $\mathbf{H}_{1:K} \in \mathbb{R}^{K\times z}$, where $K$ is the sequence length, and $z$ is the feature dimension. 

The second main component \circled{\textbf{2}} is an aggregator that reduces the length of $\mathbf{H}_{1:K}$. 
In the simplest case, this component applies mean pooling along the time dimension, resulting in a pooled representation $\mathbf{R} \in \mathbb{R}^z$ that encapsulates the global context of the audio. 
While the aggregator could also be parameterized by a neural network, we found that computing the average over the sequence yields similar results to more complex approaches (cf. Section~\ref{ssec:main_results}). 
% The aggregator component provides a summarized, global representation of the audio sequence, providing a different view over the data. 
The aggregation component generates a summarized global representation of the audio sequence, providing an alternative perspective on the data. 
This approach is inspired by prior works \cite{wagner2024multimodal,palaskar2024multimodallargelanguagemodels}, which also use aggregated representations for similar tasks. 
However, unlike models that rely exclusively on aggregated representations, our architecture retains the flexibility to handle tasks like ASR, which require detailed sequence-level information. 
Therefore, the aggregated sequence and the original sequence are concatenated $\mathbf{H}_{1:K+1} = [\mathbf{R}; \mathbf{H}_{1:K} ] \in \mathbb{R}^{(K+1) \times z}$ in the next step. 

The combined representations are then either directly passed to the LLM or further processed by the third main component \circled{\textbf{3}}, which consists of a gating function $\mathcal{G}$ similar to \cite{bondarenko2023outlier} that dynamically modulates the contribution of individual audio features. 
% now having full dimensions $B \times z \times (K + 1)$, are then passed to the LLM. 
$\mathcal{G}$ is a small neural network, parameterized by a linear layer followed by a sigmoid activation function that is jointly learned with the rest of the model. 
It generates a gating signal $\mathcal{G}(\mathbf{H}_{1:K+1})$, which is applied to $\mathbf{H}_{1:K+1}$ via the Hadamard product.
The gating network adjusts the extent to which individual audio representations influence the final latent representations. 
This allows the model to either rely heavily on or diminish the impact of specific audio features, depending on the task and input. 

The fourth main component \circled{\textbf{4}} is a decoder-only LLM, which processes the audio representations jointly with any other input text tokens, such as prompts, to generate the final output for the five target tasks. 
In the forward pass, the audio placeholder token embeddings are replaced by the representations obtained with the audio encoder. 
The combined audio/text representations are then processed through the $V$ transformer blocks of the LLM. 

We employ low-rank adaptation (LoRA) \cite{hu2022lora} to jointly optimize the audio encoder and the LLM. 
LoRA is a technique that enables model specialization by optimizing smaller adapter modules consisting of low-rank decomposition matrices, while keeping the pretrained weights of the underlying model unchanged. 
% In LoRA, the pretrained weights of the base model are kept constant and smaller adapter modules, which are inserted within transformer blocks, are optimized instead. 
On a high level, LoRA finetuning is used to introduce the new tasks related to the virtual assistant and to capture the overall domain shift (e.g. acoustic conditions). 
We chose LoRA over full finetuning due to its effectiveness on the DDSD task \cite{wagner2024multimodal,palaskar2024multimodallargelanguagemodels} and its practical benefits. 
Specifically, LoRA improves inference efficiency by loading the underlying model weights into memory once, allowing them to be shared across applications, while smaller adapter weights can be quickly switched to support different use cases.

Components \circled{\textbf{2}} and \circled{\textbf{3}} are optional, i.e., the model can be configured with or without sequence aggregation and gating. 

% Both the Qwen LLM and the audio encoder consist of multiple transformer blocks. 
% The adapters are composed of matrices $A \in \mathbb{R}^{m \times r}$ and $B \in \mathbb{R}^{r \times n}$. 
% The of the transformer blocks (denoted as $W$) remain frozen during the task-specific training. 
% The adapters modify the frozen weights in an incremental update step via $W' = W + BA$. 
% This enables the model to adjust its parameters for specific tasks without altering the pretrained model weights.

% \subsection{Frame Stacking}
% In Prompting Large Language Models with Speech Recognition Abilities a frame stacking approach is used to reduce the number of embeddings presented to the LLM. 
% The approach leads to For example, if n consecutive 512-D embeddings are stacked, we would obtain embeddings of dimensionality $n \times 512$. 

% Gated Att:
% self.alpha = nn.Linear(self.embed_dim, self.num_heads, bias=True)
% self.gate_fn = torch.sigmoid
% self.gate_scaling_factor = 1
% Forward:
    % x = hidden_states  # (B, T, d_model)
    % alpha = self.alpha(x)  # (B, T, H)
    % gate = self.gate_fn(alpha)
    % gate = gate.permute(0, 2, 1).contiguous()  # (B, H, T)
    % gate = gate.unsqueeze(3)  # (B, H, T, 1)
    % attn_output *= gate * self.gate_scaling_factor

\subsection{Training Objective}
For a training set containing $D$ log-Mel spectrograms, textual context (e.g. prompts), and task labels $\{(\boldsymbol{x}_{1:T}^{(d)}, \boldsymbol{t}_{1:N}^{(d)}, \boldsymbol{l}_{1:M}^{(d)})\}_{d=1}^D$, the training objective for the parameter set $\theta$ is the autoregressive prediction of the next token $y_i$, given the previously predicted tokens $\boldsymbol{y}_{<i}$, the textual
context $\boldsymbol{t}_{1:N}$ and the log-Mel spectrograms $\boldsymbol{x}_{1:T}$ using the cross-entropy objective:
\vspace{-3mm}
\begin{equation*}
    \mathcal{L}_{\theta} =-\sum_{i=1}^{M+N} p_{\theta}\left(y_i \mid \boldsymbol{y}_{<i}, \boldsymbol{t}_{1:N}, \boldsymbol{x}_{1:T}\right).
    \vspace{-3mm}
\end{equation*}
During inference, we use greedy search to generate a token sequence for a predefined maximum number of steps or until the \texttt{<|endoftext|>} token is encountered. 
Subsequently, class labels are extracted from the scores associated with the generated sequence. 
For the DDSD and VT detection tasks we consider the score $p_\theta\left(Y=yes \mid c \right)$ directly after the special tokens \texttt{<|DD|>} and \texttt{<|VT|>} are generated. 
$Y$ is a discrete random variable that can realize one of $k$ tokens  $y_1, ... , y_k$ from the vocabulary $\mathcal{V}$ and $p_\theta\left(Y=yes \mid c \right) + p_\theta\left(Y=no \mid c \right) \simeq 1$.  
The full context $c$ is given by the audio and textual context, as well as any previously predicted tokens including the special tokens indicating the current task, i.e., $c = ( \boldsymbol{y}_{<i}, \boldsymbol{t}_{1:N}, \boldsymbol{x}_{1:L} )$. 
We limit the generation process to a maximum of 256 new tokens in our experiments.
\subsection{Prompting for Tasks}
%%%%%%%%%%%%%%%%%%%% Full table - all prompts %%%%%%%%%%%%%%%%%%%%%%%
% \begin{table}[t]
% \setlength{\tabcolsep}{1pt}
% \caption{Prompts} 
% \centering
% \def\arraystretch{1.1} \small
% \begin{tabular}{|c|l|}
% \hline
% \textbf{Task} & \textbf{Prompt} \\
% \hline
% ASR & what does the person say? \\
% \hline
%  VT & does this query contain the trigger phrase? \\
% \hline
% DDSD & \makecell[l]{is this query directed \\ towards a virtual assistant?} \\
% \hline
% DA & what type of dialog act is this? \\
% \hline
% ODLD & \makecell[l]{is this ASR transcript intended \\ for a virtual assistant or not?} \\
% \hline
% ASR + DA & \makecell[l]{what does the person say and \\ what type of dialog act is this?}  \\
% \hline
% ASR + VT & \makecell[l]{what does the person say and \\ does this query contain the trigger phrase?} \\
% \hline
% ASR + DDSD & \makecell[l]{what does the person say and \\ is this query directed towards a virtual assistant?} \\
% \hline
% ASR + DA & \makecell[l]{what does the person say and \\ what type of dialog act is this?} \\
% \hline
% VT + DA & \makecell[l]{does this query contain the trigger phrase and \\ what type of dialog act is this?} \\
% \hline
% \end{tabular}
% \label{tab:prompts}
% \end{table}
%%%%%%%%%%%%%%%%%%%% End Full table - all prompts %%%%%%%%%%%%%%%%%%%%%%%

\begin{table}[t]
\setlength{\tabcolsep}{10pt}
\caption{Prompts used to distinguish between tasks.} 
\vspace{-2mm}
%\resizebox{0.49\textwidth}{!}{
\scriptsize
\centering
\def\arraystretch{1.1}
\begin{tabular}{|c|c|l|}
\hline
\textbf{\#} & \textbf{Task} & \textbf{Prompt} \\
\hline
1 & ASR & \makecell[l]{What does the person say?}  \\
\hline
2 & ASR + DA & \makecell[l]{What does the person say and \\ what type of dialog act is this?}  \\
\hline
3 & ASR + VT & \makecell[l]{What does the person say and \\ does this query contain the trigger phrase?} \\
\hline
4 & ASR + DDSD & \makecell[l]{What does the person say and \\ is this query directed towards a virtual assistant?} \\
\hline
5 & ASR + DA & \makecell[l]{What does the person say and \\ what type of dialog act is this?} \\
\hline
6 & VT + DA & \makecell[l]{Does this query contain the trigger phrase and \\ what type of dialog act is this?} \\
\hline
\end{tabular}
%}
\label{tab:prompts}
\vspace{-5mm}
\end{table}

%We distinguish between different tasks via prompts that are passed as textual input to the model at training and inference time.  
We differentiate tasks by using prompts, which are provided as textual input to the model during both training and inference. 
The prompts for each task are listed in Table~\ref{tab:prompts}. 
% The first five rows are prompts for individual tasks, where the model considers each task in isolation. 
%The second half of the table shows combined prompts, in which the model is asked to first perform one task (e.g. ASR) followed by another (e.g. VT detection). 
The model is trained to first perform one task (e.g. ASR) followed by another (e.g. VT detection). 
In this combined approach, the system first transcribes the audio input, generating a text representation of the spoken words and then analyzes the transcript alongside the acoustic cues to make a decision (e.g. determining whether the trigger phrase is present). 

By first converting speech to text, we aim to encourage the system to also make use of the textual representations to better understand the context and content of the query. 
For example, in an ideal case ASR provides a clear and correct text-based context, helping to distinguish between similar-sounding words and non-vocative uses of the trigger word, consequently making a better VT decision. 
% Non-vocative refers to instances where the trigger word is mentioned in a different context not meant to trigger the device. 
In noisy conditions, ASR can attempt to filter out background noise and focus on the spoken words, which may enhance the model's robustness to challenging acoustic environments.
Our hope is that by integrating ASR into the VT, DDSD, and DA process helps addressing key sources of errors by providing a clearer, text-based representation of the audio input, enhancing the system's ability to accurately identify trigger words, improving overall performance and reliability. 
Furthermore, joint ASR training should attribute more weight to salient information, such as trigger words compared to independently trained ASR that will weigh trigger words equally as inconsequential propositions. 

% \begin{itemize}
%     \item Use frame stacking as in Prompting Large Language Models with Speech
% Recognition Abilities https://arxiv.org/pdf/2307.11795 to reduce the number of Whisper embeddings as an alternative to QFormer
%     \item Learn to ignore whisper encodings in favor of aggregated embedding similar to gated attention?
% \end{itemize}

\section{Experiments}
\subsection{Modeling Details}
We use the same training procedure for all SELMA variants in our experiments. 
Each model is trained on $16 \times$ A100 40GB GPUs for 350k steps with an effective batch size of 256. 
For optimization, we use AdamW \cite{loshchilov2018decoupled} ($\lambda = 10^{-4}, \epsilon = 10^{-8}, \beta_1 = 0.99, \beta_2 = 0.999$) with an initial learning rate of $2 \times 10^{-4}$, a linear schedule and a warm-up phase of 10\% of total training steps. 
% ["attn.query", "attn.value", "attn.c_attn"]
We attach LoRA modules to the query and value matrices of both the audio encoder and the LLM part of the model.  
The overall system has 5.5M trainable parameters ($\sim$0.84\% of the overall model). 
We use gradient clipping with a maximum $L_2$ norm of 1. 
We set $r = 8$ and the scaling factor for adjusting the magnitude of the adaption $\alpha = 32$ in all our experiments. 
The LoRA modules are optimized with a dropout probability of 10\%. 
\subsection{Data}
We employ VT detection training data similar to \cite{rudovic23sdsd} and \cite{higuchi2024multichannelvoicetriggerdetection}, which comprises clean audio that has been further augmented with various types of noise and room impulse responses. 
The training set contains $\sim$7M queries with an average duration of $\sim$1.3 seconds. 
We exclude $\sim$121k randomly sampled queries from the training set to serve as the validation set. 
The VT detection task is evaluated on an in-house test set that is also similar to \cite{rudovic23sdsd} and \cite{higuchi2024multichannelvoicetriggerdetection}. 
The test data is comprised of $\sim$130k queries with an average duration of 1.4 seconds. 

The training data for DDSD are updated versions of the sets used in \cite{wagner2024multimodal} comprising a total of $\sim$237k queries (average duration 4.9 seconds) and an additional $\sim$3.8M sentences of text-only data without corresponding audio. 
We add the text-only data for regularization,  to help the model not to rely exclusively on the audio modality. 
The DDSD validation data comes from the same source as the training data and comprises $\sim$4k queries (average duration 4.8 seconds).  
% and the ODLD validation data comprises $\sim$43k sentences. 
We evaluate the DDSD task on an updated version of the in-house test from \cite{wagner2024multimodal}, which is comprised of $\sim$36k queries (average duration 3.5 seconds). 
The ASR training and validation data are a mixture of randomized and anonymized in-house corpora with a total of $\sim$232k (average duration 4.2 seconds) and $\sim$10k (average duration 4.1 seconds) queries respectively. 
The ASR test data are a randomized and anonymized in-house corpus consisting of $\sim$20k queries with an average duration of $\sim$4.2 seconds. 
The auxiliary DA classification training data uses a randomly sampled subset of the VT and DDSD training data, which has been automatically annotated using an off-the-shelf Phi-3 14B model \cite{abdin2024phi3technicalreport}. 

The VT and DDSD corpora lack ground truth transcripts. 
However, our goal is to enable the model to perform ASR followed by another task (cf. Table~\ref{tab:prompts}). 
Therefore, we first train a auxiliary model that uses the same architecture and configuration as our main model. 
We formulate the VT detection and DDSD tasks such that the model is not required to perform ASR first, i.e., we use the prompts from Table~\ref{tab:prompts} but without the ASR-related part. 
% The trained auxiliary model is then used to generate ASR transcripts (via prompt \#1 in Table~\ref{tab:prompts}) for the VT and DDSD datasets. 
% These ASR transcripts form the ground truth labels during training of the main model.  
The trained auxiliary model is subsequently utilized to generate ASR transcripts for the VT and DDSD datasets, as specified by prompt \#1 in Table~\ref{tab:prompts}. 
These ASR transcripts serve as the ground truth labels during the main model's training process.

% VT = 15
% ASR = 30
% FF = 35
% Main = 15 + 30 + 35 = 80
% DA = 15
% ODLD = 5
The training data is weighted such that the main tasks account for 80\% (15\% VT, 35\% DDSD, 30\% ASR) of the overall data during training. 
The auxiliary tasks account for the remaining 20\% of the data (5\% text-based DDSD and 15\% DA). 
We found that this weighting scheme yielded robust results across all tasks in our preliminary experiments. 

\subsection{Results and Discussion}\label{ssec:main_results}
Table~\ref{tab:exp} shows the performance of various models on the DDSD, VT detection and ASR tasks, evaluated using Equal-Error Rate (EER) and Word Error Rate (WER).
The baseline models include zero-shot results obtained with the Qwen-Audio-Chat model, ASR results using a 1.55B parameter Whisper model finetuned with LoRA on the same ASR data as the SELMA models, as well as DDSD results obtained with the systems described in \cite{wagner2024multimodal} using GPT2-XL and Qwen 7B as LLM backbones, Whisper as the audio encoder and the same DDSD training data as the SELMA models. 
Furthermore, we compare SELMA to the Unified Acoustic Detector (UAD) developed in \cite{rudovic23sdsd}, as well as the text-based Out-of-Domain Language Detector (ODLD) described in \cite{voicetrigger23}. 
% Notably, Qwen-Audio (zero-shot) exhibits high error rates for DDSD (51.7%) and VT (40.0%), while achieving a relatively moderate WER of 0.348 for ASR. Whisper (large-v2) has a WER of 0.441, but its finetuned variant significantly improves ASR performance with a WER of 0.101. GPT2-XL + Whisper and Qwen 7B + Whisper show moderate DDSD EERs of 10.00% and 10.78%, respectively. The BERT finetuned model, focusing solely on text, achieves the highest DDSD EER of 12.32% and VT EER of 2.18%, further demonstrating the limitations of using text-only models for such tasks.

The experiments under ``Changing Model Components'' and ``Removing Auxiliary Tasks'', show the performance of ``SELMA 1'', when various data- and model-related modifications are applied. 
The experiment denoted as ``SELMA 1'', which uses a concatenation of the mean pooled audio representation sequence and the sequence itself, exhibits strong performance across all tasks, with a DDSD EER of 7.78\%, VT EER of 0.12\%, and WER of 0.125. 
% In contrast, ``SELMA 2'', which uses only mean pooling, i.e., the LLM backbone relies on a single vector to represent the audio, shows higher EERs for DDSD (8.76\%) and VT (0.28\%), alongside a significant increase in WER to 0.398. 
% ``SELMA 3'', which uses only the sequence of audio representations and no pooling performs worse than ``SELMA 1'' on all three tasks. 
% ``SELMA 4'', which combines mean pooling, concatenation, and the gating network, improves the DDSD EER to 7.63\% but slightly increases the VT EER to 0.19\% and WER to 0.134. 
% ``SELMA 5'' employs a more advanced aggregation mechanism using the Q-Former model from \cite{tang2024salmonn}. 
% However, the results do not approve over the simple mean pooling employed in ``SELMA 1''. 

% \circled{\textbf{2}} is an aggregator that reduces the length of $\mathbf{H}_{1:K}$. 
% In the simplest case, this component applies mean pooling along the time dimension, resulting in a pooled representation $\mathbf{R} \in \mathbb{R}^z$
``SELMA 2'', which uses only mean pooling, i.e., only the representation $\mathbf{R}$ obtained via component \circled{\textbf{2}} is used as audio context, shows higher EERs (8.76\% for DDSD, 0.28\% for VT) and a significant WER increase to 0.398. 
``SELMA 3'', where component \circled{\textbf{2}} is disabled, i.e., no pooling is applied, also performs worse on all tasks. 
``SELMA 4'', combining mean pooling, concatenation, and the gating network (component \circled{\textbf{3}}), improves DDSD EER to 7.63\% but performs slightly worse on VT detection and ASR. 
``SELMA 5'' employs aggregation via the Q-Former model from \cite{tang2024salmonn} but does not outperform the simpler mean pooling of ``SELMA 1''.

\begin{table}[t!]
\setlength{\tabcolsep}{3pt}
\caption{Results on DDSD, VT, and ASR.} 
\vspace{-2mm}
\centering 
\def\arraystretch{1} 
\scriptsize
%\resizebox{0.49\textwidth}{!}{
\begin{tabular}{|c|l|c|c|c|}
\hline 
\multicolumn{2}{|c|}{\textbf{Experiment}} & \multicolumn{2}{c|}{\textbf{EER} ($\downarrow$) } & \textbf{WER} ($\downarrow$) \\
\multicolumn{2}{|c|}{{}} & DDSD  & VT & ASR \\
\hline
\multirow{6}{*}{Baselines} & Qwen-Audio-Chat (zero-shot) \cite{chu2023qwenaudio} & 51.7\% & 40.0\% & 0.348 \\
%& Whisper (\texttt{large-v2}) \cite{radford2022robust} & - & - & 0.441\\
& Whisper finetuned (\texttt{large-v2}, LoRA) & - & - & 0.101 \\
& GPT2-XL + Whisper \cite{wagner2024multimodal} & 10.00\% & - & - \\
& Qwen 7B + Whisper \cite{wagner2024multimodal} & 10.78\% & - & - \\
& UAD \cite{rudovic23sdsd} & 11.07\% & 0.33\% & - \\
& ODLD \cite{voicetrigger23} & 12.32\% & 2.18\% & - \\
\hline
\hline
SELMA 1 & Mean pooling + sequence & 7.78\% & 0.12\% & 0.125 \\
\hline
\multicolumn{5}{|c|}{\textbf{Changing Model Components}} \\ 
\hline
SELMA 2 &  Mean pooling only & 8.76\% & 0.28\% & 0.398 \\
SELMA 3 &  Sequence only & 8.06\% & 0.17\% & 0.129 \\
SELMA 4 &  Mean pooling + sequence + gating & 7.63\% & 0.19\% & 0.134 \\
SELMA 5 &  Q-Former \cite{tang2024salmonn} + concatenate & 7.94\% & 0.19\% & 0.127 \\
\hline
\multicolumn{5}{|c|}{\textbf{Removing Auxiliary Tasks}} \\ 
\hline
SELMA 6 &  No text-only DDSD & 7.78\% & 0.24\% & 0.131 \\
SELMA 7 &  No text-only DDSD, no DA & 7.50\% & 0.18\% & 0.125 \\
SELMA 8 &  No text-only DDSD, no DA, no ASR & 7.45\% & 0.22\% & 0.216 \\
% - ODLD - DA - ASR; seq only & xxx\% & xxx\% & xxx\% \\
\hline
\end{tabular}
%}
\label{tab:exp}
\vspace{-5mm}
\end{table}

% In the context of removing auxiliary tasks, the ``SELMA 6'' model, which excludes the text-based DDSD task and training data, maintains the same DDSD EER (7.78\%) as ``SELMA 1'' but shows a degradation in VT EER (0.24\%) and WER (0.131). 
% The ``SELMA 7'' model, which excludes both text-based DDSD and DA, shows a small improvement in DDSD EER (7.50\%) and a WER (0.125) comparable to ``SELMA 1'', but an EER degradation from 0.12\% to 0.18\% on the VT task. 
% Finally, the ``SELMA 8'' model, which also excludes the ASR task (cf. prompt \#1 in Table~\ref{tab:prompts}) from training shows a notable increase in WER to 0.216. 
% Note that the ``SELMA 8'' model is still trained to perform ASR via the combined prompts \#3 and \#4 from Table~\ref{tab:prompts}. 
% However, it is not tasked to perform ASR as a standalone task anymore, which also means that it is no longer trained on high-quality human-annotated ground truth transcripts. 
``SELMA 6'', without the text-based DDSD task and training data, maintains DDSD EER (7.78\%) but worsens VT detection EER (0.24\%) and WER (0.131). 
``SELMA 7'', excluding text-based DDSD and DA classification, slightly improves DDSD EER (7.50\%) and maintains the WER (0.125) but degrades VT detection EER to 0.18\%. 
``SELMA 8'', additionally excluding ASR as a standalone task, shows a notable WER increase to 0.216 due to the lack of high-quality human-annotated transcripts.

Overall, the results show that the combination of mean pooling and the original sequence used in ``SELMA 1'' are crucial for achieving robust performance across all tasks, indicating that the model can extract more information from the full sequence, while providing an additional global view of the sequence is beneficial for DDSD and VT detection. 

\subsection{DET Curves}
\begin{figure}[t]
\centering
\includegraphics[width=0.46\textwidth]{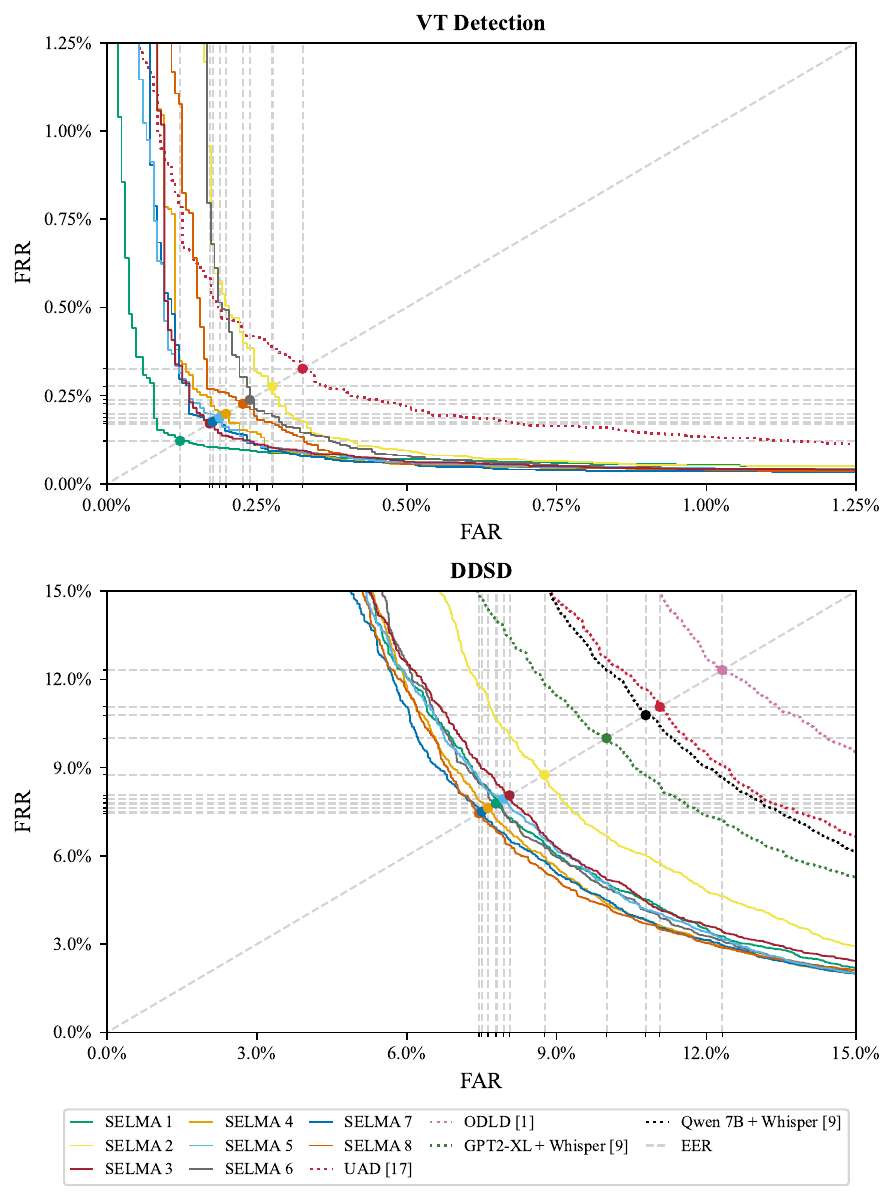}
\vspace{-2mm}
\caption{DET curves for VT detection and DDSD experiments from Table~\ref{tab:exp}. The False Accept Rate (FAR) represents either unintended queries or queries without the trigger phrase that were falsely classified as intended/containing the trigger phrase and the False Reject Rate (FRR) represents either intended queries or queries containing the trigger phrase that were falsely classified as unintended/not containing the trigger phrase. 
The markers on each curve indicate the EER. Dotted lines represent baselines and solid lines represent our approach.}
\label{fig:eer}
\vspace{-5mm}
\end{figure}

%%%%%%%%%%%%% TODO: Describe other experiments a bit more %%%%%%%%%%%%%%%%%%%%%%%%%%%%%%
Figure~\ref{fig:eer} shows the Detection Error Trade-off (DET) curves for the VT detection (top) and DDSD (bottom) experiments. 
The EERs, marked by the intersection of the DET curve with the diagonal line correspond to the results presented in Table~\ref{tab:exp}. 
The SELMA models (solid lines) outperform the UAD baselines (dotted line) on the VT detection task, achieving lower FAR and FRR across a wide range of thresholds. 
The ``SELMA 1'' model (solid green line) outperforms the UAD baseline across all operating points. 
In the DDSD plot, the SELMA models demonstrate better performance than the baselines across all operating points.
The SELMA models exhibit tightly grouped curves, showing little impact on performance across the various configurations with the only exception being ``SELMA 2'', where the model relies on a mean pooled audio representation only. 
%In the DDSD plot (bottom), SELMA models also achieve better performance, with lower EER values compared to baselines such as UAD and GPT2-XL + Whisper. The SELMA models exhibit tightly grouped curves, demonstrating consistent performance across the various configurations. Notably, SELMA 4 and SELMA 8 show the best trade-offs, with EERs around 7-8%, indicating their robustness in detecting device-directed speech. In contrast, the baselines show higher EERs, with wider curves, illustrating their lower accuracy and higher variability in handling DDSD tasks.
Figure~\ref{fig:eer} confirms the findings from Table~\ref{tab:exp}, demonstrating that our model maintains a more favorable trade-off between false accepts and false rejects in both VT detection and DDSD tasks. 

\vspace{-0.2mm}
\section{Conclusions}
% \textbf{S}peech-\textbf{E}nabled \textbf{L}anguage \textbf{M}odel for Virtual \textbf{A}ssistant Interactions
% We presented SELMA, a speech-enabled language model for virtual assistant interactions capable of performing multiple tasks related to processing user input for virtual assistants. 
% SELMA integrates audio and text modalities and understands various tasks related to interactions with virtual assistants including speech recognition, voice trigger detection, and device-directed speech detection. 
% We explored different training approaches and model architectures to handle multiple tasks simultaneously while maintaining high performance across all tasks. 
We presented SELMA, a Speech-Enabled Language Model for virtual Assistant interactions. 
By integrating multiple tasks such as automatic speech recognition, voice trigger detection, and device-directed speech detection within a single system, as well as by performing end-to-end training of the audio encoder and LLM backbone, SELMA achieves superior performance compared to systems optimized for individual tasks. 
SELMA not only reduces the complexity of a typical pipeline for user input processing in virtual assistants significantly, but also enhances generalization and robustness across various tasks along the pipeline. 
Future work will explore the integration of downstream NLU components, covering the entire user input processing pipeline with a single unified model. 

\newpage
\bibliographystyle{IEEEbib}
\footnotesize{
    \bibliography{refs}
}
\end{document}